\documentclass[aps,prl,twocolumn,superscriptaddress,preprintnumbers]{revtex4-2}
\pdfoutput=1
\usepackage{amsmath}
\usepackage{graphicx}
\usepackage{bbm}
\usepackage{amssymb}
\usepackage{enumitem}
\usepackage{booktabs}
\usepackage{verbatim} 
\usepackage{soul}
\usepackage{mathrsfs}
\usepackage[linktocpage=true,
colorlinks=true,
urlcolor=blue,
anchorcolor=blue,
citecolor=blue,
filecolor=blue,
linkcolor=blue,
menucolor=blue,
]{hyperref}

\usepackage{color}

\begin{document}

\title{
No manifest T-duality at order $\alpha'^3$}

\author{Steven Weilong Hsia}
\email{hsiasteven000@gmail.com}
\affiliation{Department of Theoretical Physics and Astrophysics, Faculty of Science, Masaryk University, Kotlářská 2, CS-61137 Brno, Czechia}

\author{Ahmed Rakin Kamal}
\email{ahmedrakinkamaltunok@gmail.com}
\affiliation{Department of Theoretical Physics and Astrophysics, Faculty of Science, Masaryk University, Kotlářská 2, CS-61137 Brno, Czechia}
\affiliation{Department of Mathematics and Natural Sciences, BRAC University, 66 Mohakhali, Dhaka 1212, Bangladesh}

\author{Linus Wulff}
\email{wulff@physics.muni.cz}
\affiliation{Department of Theoretical Physics and Astrophysics, Faculty of Science, Masaryk University, Kotlářská 2, CS-61137 Brno, Czechia}

\begin{abstract}
\noindent When reduced from $10$ to $10-d$ dimensions tree-level string theory exhibits an $O(d,d)$ symmetry. This symmetry, which is closely related to T-duality, appears only after certain field redefinitions. We find a simple form for a subset of these redefinitions at order $\alpha'^3$ and show that they cannot be lifted to ten dimensions. This is inconsistent with ``manifestly T-duality invariant'' approaches such as generalized geometry (in the uncompactified setting). Such formulations therefore seem not to be the correct language to describe string theory.
\end{abstract}

\maketitle

\section{Introduction}\label{sec:intro}
In string theory, the supergravity action is accompanied by an infinite series of higher derivative corrections. The corrections are organized in powers of the string coupling $g_s$ and inverse string tension $\alpha'$. In this work we restrict ourselves to tree-level string theory, i.e. lowest order in $g_s$. We still have an infinite series of $\alpha'$-corrections. The first such correction is at first order in $\alpha'$ in the case of type I, heterotic or bosonic strings and third order in the type II case. The former correction involves the square of the Riemann tensor \cite{Metsaev:1987bc,Metsaev:1987zx}, while the latter, which is present for all string theories, involves the fourth power of the Riemann tensor with a coefficient proportional to $\zeta(3)$ \cite{Gross:1986iv, Grisaru:1986vi,Freeman:1986zh,Cai:1986sa,Gross:1986mw,Jack:1988sw}.

The terms involving only the metric have been found from 4pt and 5pt tree-level string graviton amplitudes and have been known for a long time. However, to find the complete correction involving all the massless fields at order $\alpha'^3$ would require going up to 8pt and matching the amplitude to an ansatz for the effective action with $\mathcal O(1000)$ terms \cite{Garousi:2020mqn}.

The T-duality symmetry of string theory offers a possible short-cut (another possibility is to use supersymmetry, e.g. \cite{Ozkan:2024euj}). Restricting to the NSNS sector fields ($G,B,\Phi$) common to all string theories, T-duality on a circle turns out to completely fix the correction at order $\alpha'$ and $\alpha'^3$, up to the overall coefficient \cite{Razaghian:2017okr,Garousi:2020gio}. This approach still involves writing a very large general ansatz for the action and had to be carried out with the aid of a computer, leading to a complicated final answer. But this approach can be considerably simplified by realizing that it is enough to look only at a small subset of the terms arising in the dimensional reduction, as we will now explain.

In general, reducing string theory from $10$ to $10-d$ dimensions leads to the appearance of an $O(d,d)$ symmetry \cite{Meissner:1991zj, Maharana:1992my}, which is closely connected with T-duality. This symmetry persists to all orders in $\alpha'$ for the massless fields \cite{Sen:1991zi}, but is broken by the massive modes to a discrete subgroup. It therefore applies to the tree-level low-energy effective action. Focusing on the KK vectors (displaying only the internal index $n'=10-d,\cdots,9$)
\begin{equation}\label{eq:AAhat}
    A_{n'} = -\tfrac12 (A^{(1)}_{n'} + A^{(2)}_{n'}), \qquad \hat{A}_{n'} = \tfrac12 (A^{(1)}_{n'} - A^{(2)}_{n'})\,,
\end{equation}
where $A^{(1)}$ comes from the metric and $A^{(2)}$ from the $B$-field, the subgroup $O(d)\times O(d)\subset O(d,d)$ acts by independently rotating $A$ and $\hat A$. A necessary condition for $O(d,d)$ symmetry is therefore that any terms involving the $O(d,d)$ violating contraction $A\cdot\hat A\equiv A_{n'}\hat A^{n'}$ must cancel out in the reduced action. This condition was used in \cite{Wulff:2021fhr,Wulff:2024ips, Wulff:2024mgu} to fix the form of the corrections up to order $\alpha'^3$ (in the latter case modulo terms of sixth order and higher in fields).

Actually, the $O(d,d)$ violating terms in the reduced action cancel only up to terms proportional to the lowest order (SUGRA) equations of motion. These terms can then be removed by field redefinitions (up to terms of yet higher order in $\alpha'$). Here we analyze the form of the required field redefinitions (or, equivalently, modifications of the $O(d,d)$ transformations) and show that at order $\alpha'^3$ they cannot be uplifted to ten dimensions. Said another way: there is no (local) choice of fields, a.k.a. scheme, in $D=10$ which, upon dimensional reduction, leads to a reduced action which is directly $O(d,d)$ invariant without further field redefinitions (or modifications of the transformations).

This is in direct conflict with approaches like generalized geometry \cite{Hitchin:2003cxu,Gualtieri:2003dx,Grana:2008yw} (and the closely related double/exceptional field theories (DFT/ExFT), reviewed in \cite{Aldazabal:2013sca,Berman:2013eva,Hohm:2013bwa,Hohm:2019bba}), where the $O(d,d)$ symmetry is present from the outset. Therefore, contrary to what is often assumed, such ``manifestly T-duality invariant'' formulations seem not to be the correct language to describe string theory. This observation is consistent with a previously found obstruction in DFT \cite{Hronek:2020xxi}. Note that this problem is there only when describing the general uncompactified theory, there is no problem when restricting to backgrounds with $d$ abelian isometries.

\section{Order $\alpha'$}\label{sec:alpha prime order}
As a warm-up, let us consider the order $\alpha'$ correction for the bosonic string. Ignoring the overall coefficient it can be written \cite{Wulff:2024ips}
\begin{equation}
    L_{\alpha'} = \mathcal R_{abcd}\mathcal R^{abcd}-H^2_{abcd}\mathcal R^{abcd}-\tfrac13H^2_{abcd}(H^2)^{acbd}-H^2_{ab}\mathbb G^{ab}
\end{equation}
where we defined $H^2_{abcd}\equiv H_{abe}H^e{}_{cd}$, $H^2_{ab}\equiv H_{acd}H^{cd}{}_b$ and introduced the torsionful Riemann tensor
\begin{equation}\label{eq:Rcal}
\mathcal R_{abcd}=R_{abcd}-\nabla_{[a}H_{b]cd}+\tfrac12H^2_{a[cd]b}\,.
\end{equation}
The last term, proportional to the metric equation of motion
\begin{equation}
\mathbb G_{ab}\equiv R_{ab}+2\nabla_a\nabla_b\Phi-\tfrac14H^2_{ab}\,,
\end{equation}
has been added for convenience (it can be removed by a field redefinition). KK reduction (see appendix) gives, in addition to $O(d,d)$ invariant terms, the following $O(d,d)$ \emph{violating} term quadratic in the KK vectors (when there is no risk of confusion we use the same notation for indices in $D=10$ and $D=10-d$)
\begin{equation}\label{eq:term00}
    L_{\alpha'}\rightarrow -4\hat{F}^{ac}\cdot F^b{}_c(\mathbb {G}_{ab} + \tfrac12 \mathbb{B}_{ab})\,,
\end{equation}
where $F$($\hat F$) are the field strengths of the KK vectors (\ref{eq:AAhat}) and the equation of motion for the $B$-field is 
\begin{equation}
\mathbb B_{ab}=(\nabla^c-2\partial^c\Phi)H_{abc}\,.
\end{equation}
Since it is proportional to the equations of motion this term can be removed by a field redefinition. This is in the reduced theory, but it is interesting to ask whether it can be removed already from the start. For this to happen one needs a term in $26$ dimensions which, upon KK reduction, cancels this term. It is not hard to find such a term in the present case, but if we could not find it, how would we check for an obstruction? 

The strategy we will use is to first linearize the term. This leads to several terms and we pick one of them. In the present case the most interesting term is one coming from the $B$ e.o.m., namely
\begin{equation}\label{eq:term0}
    -2\partial^k \hat A^m\cdot\partial^l A_m\partial^2 B_{kl}\,.
\end{equation}
To see if this term can be uplifted to $26$ dimensions one needs to write all terms which can contribute to this term upon KK reduction. At the linearized level the KK reduction ansatz is (we are ignoring any terms involving the KK scalars since we don't need these terms)
\begin{equation}\label{eq:red1}
\underline{h}_{\underline{mn}} = \begin{pmatrix}
        h_{mn} 
        & A^{(1)}_{n'm} \\
        A^{(1)}_{m'n} & 0
    \end{pmatrix}\,, \quad \underline{B}_{\underline{mn}} = \begin{pmatrix}
        B_{mn} & A^{(2)}_{n'm} \\
        -A^{(2)}_{m'n} & 0
    \end{pmatrix}\,.
\end{equation}
The index structure of the term remains basically unchanged in the reduction, except that one pair of indices become internal, therefore terms of different index structure will not mix. Using also the fact that the action is symmetric under $B\rightarrow-B$ one finds that it is not possible to construct a term out of the linearized metric and $B$-field which can cancel the term (\ref{eq:term0}). However, if we take instead of the metric the linearized vielbein, $e\sim1+f$, with KK reduction ansatz
\begin{equation} \label{eq:redf}
    \underline{f}_{\underline{mn}} = 
    \begin{pmatrix}
        f_{mn} & A^{(1)}_{n'm} \\ 
        0 & 0
    \end{pmatrix},
\end{equation}
then the term
\begin{equation}
    \partial^kf^{mn}\partial^lB_{mn}\partial^2 B_{kl}
\end{equation}
does the job. This term would simply vanish if $f$ were symmetric, which shows the problem with using only the metric. This observation is consistent with the fact that the full nonlinear term in (\ref{eq:term00}) can be uplifted using $H_{abc}$ and the spin connection \cite{Marques:2015vua,Wulff:2024ips}. This example highlights the fact that the answer can depend on the field content (and gauge symmetries) we assume in the higher dimensional theory.

\section{Obstruction at order $\alpha'^3$}
The approach of requiring the $O(d,d)$ violating terms quadratic in the KK vectors to cancel was first implemented in \cite{Wulff:2021fhr} for the $\alpha'^3$ correction. The calculation was revisited, keeping track also of the dilaton and terms proportional to the equations of motion in \cite{Wulff:2024mgu}. The e.o.m. terms were kept track of only to first order in $H$, which is all we will consider here. The expressions are quite long, but it turns out that almost all terms can be canceled by adding suitable e.o.m. terms to the $D=10$ Lagrangian (for details see the appendix). This leads to the following $D=10$ Lagrangian
\begin{equation}\label{eq:L3}
\begin{split}
L_{\alpha'^3}
&=
\tfrac{1}{4!}t_8t_8\mathcal R^4
-\tfrac{1}{2(3!)}t_8t_8H^2\mathcal R^3
+\tfrac{1}{4(4!)}\varepsilon_8\varepsilon_8\hat{\mathcal R}^4
\\
&\quad
+\tfrac{1}{2(3!)^3}\varepsilon_9\varepsilon_9H^2\hat{\mathcal R}^3
+\tfrac{1}{2(3!)^3}\varepsilon_9\varepsilon_9[H^2]\hat{\mathcal R}^3
\\
&\quad
+2\varepsilon_4H^2\varepsilon_4\mathcal R^2\mathcal R
+32H_a{}^{bc}H^{def}\mathcal R_{gh[de}\mathcal R^{ag}{}_{bf]}\hat{\mathcal R}^h{}_c
\\
&\quad
+32H^a{}_{bc}H_{def}\mathcal R^{[de|gh|}\mathcal R^{bf]}{}_{ag}\hat{\mathcal R}^c{}_h
\\
&\quad
-\tfrac{4!}{2}H^{abc}H^{def}\mathcal R^{gh}{}_{[ab}\mathcal R_{de]gh}\hat{\mathcal R}_{cf}
+\mathcal O(H^4)\,,
\end{split}
\end{equation}
where $\hat{\mathcal R}$ means that the Ricci tensor and scalar are completed to the e.o.m., e.g. $\hat{\mathcal R}_{ab}\equiv \mathcal R_{ab}+2\nabla^{(-)}_a\partial_b\Phi$, see the appendix for further details. Up to the e.o.m. terms (equivalently field redefinitions) this agrees with previous expressions in the literature \cite{Liu:2019ses,Liu:2022bfg,Garousi:2022ghs}. The point of this more complicated Lagrangian is that it leads only to the following six e.o.m. terms to linear order in $H$ (here and in the following we drop the dots, e.g. $\hat FF=\hat F\cdot F$)
\begin{equation}\label{eq:tocancel}
\begin{split}
& 32 \hat{F}_{a b}F_{e f} \mathcal{R}_g{ }^{a e c} \mathcal{R}^{g f b d} \mathbb{B}_{c d}
-32 \hat{F}^{a c}F^{b d} H_{k c d} \nabla^k \mathcal{R}_{a e f b} \mathbb{G}^{e f}
\\
&
+64 \hat{F}^{b k}F_{e f} \nabla^e H_{k c d} \mathcal{R}_{b g}{ }^{f c} \mathbb{G}^{d g}
+16 \hat{F}^{a b}F_{c d} \nabla_{(a} H_{f)}{}^{ c e} \mathcal{R}^{df}{ }_{eb} \mathbb{D}
\\
&
+64 \hat{F}^{b k}\nabla_e F^{f c} H_{k c d} \mathcal{R}_b{ }^{g e d} \mathbb{G}_{f g}
-16 \hat{F}^{a c}\nabla^g F^{b d} H_{k c d} \mathcal{R}_{a g b}{ }^k \mathbb{D} 
\\
& +(\mbox{e.o.m.})^2\mbox{-terms}+\mathcal{O}\left(H^2\right)+(H \rightarrow-H, \hat{F} \leftrightarrow F)\,,
\end{split}
\end{equation}
which now involves also the dilaton e.o.m.
\begin{equation}
\mathbb D
\equiv
R+4\nabla^d\partial_d\Phi-4\partial^d\Phi\partial_d\Phi-\tfrac{1}{12}H^2\,.
\end{equation}
Above we have dropped terms of order $H^2$ and terms quadratic in the equations of motion. The addition of the same terms with the sign of $H$ flipped and $F$ and $\hat F$ exchanged implements the ``$B$-parity'' symmetry under $B\rightarrow-B$ of the tree-level NSNS sector.

These six terms all have a different structure and can be analyzed independently. One can show that none of them can be uplifted to ten dimensions. We will illustrate the problem by looking at the first term. As before we linearize and pick one specific contribution. In this case we take \footnote{
One might worry about the possibility of adding total derivative terms. However, after linearizing the terms in (\ref{eq:tocancel}), it is not hard to convince oneself that this does not change the analysis in any essential way. In fact, since all terms vanish on-shell, it is enough to consider total derivative terms which vanish on-shell, since otherwise we would introduce new terms which don't vanish on-shell. Consider the term at hand. Since it has no symmetry between the fields, it is easy to see that the term whose derivative we add to cancel this term must itself be proportional to $\partial^2B$. But adding such a term generates terms involving additional derivatives of $\partial^2B$. Since the structure of the term is still the same as in (\ref{eq:term1}), except that some derivatives are moved to the last factor, the rest of the argument goes through in the same way, since the placement of the derivatives in the term does not affect the analysis (as long as it does not lead to any symmetry between the factors)
}
\begin{equation} \label{eq:term1}
\partial_i\hat A_j\partial_kA_l\partial_p\partial^kh^{im}\partial^p\partial^jh^{ln}\partial^2 B_{mn}\,.
\end{equation}
The most general term in ten dimensions which can contribute to this term upon reduction is
\begin{equation}\label{eq:Term1}
    C_{\alpha_1 \alpha_2 \alpha_3 \alpha_4 \alpha_5}\partial^iX^{\alpha_1}_{jq} \partial_kX^{\alpha_2,lq} \partial_p\partial^kX^{\alpha_3}_{im} \partial^p\partial^jX^{\alpha_4}_{ln} \partial^2 X^{\alpha_5,mn}\,,
\end{equation}
with
\begin{equation}
X^\alpha_{mn} = \{h_{mn}, B_{mn}\}\quad\mbox{for}\quad\alpha = 1,2\,.
\end{equation}
To cancel the term in question (and its $B$-parity partner, which we have not written here explicitly) we need to consider
\begin{equation}\label{eq:Term2}
    C_{\alpha_1 \alpha_2 112}\partial^iX^{\alpha_1}_{jq} \partial_kX^{\alpha_2,lq} \partial_p \partial^kh_{im} \partial^p\partial^jh_{ln} \partial^2 B^{mn}\,,
\end{equation}
leaving four unfixed coefficients $C_{\alpha_1 \alpha_2 112}$. But $B$-parity requires $C_{11112} = C_{22112} = 0$ (odd number of $2$'s means odd power of $B$). So, from the above terms we get, using the reduction ansatz in (\ref{eq:red1}), (from the $q$-index being internal)
\begin{eqnarray}
&(-C_{12112} + C_{21112})\partial_i\hat A_j\partial_kA_l\partial_p\partial^kh^{im}\partial^p \partial^jh^{ln} \partial^2 B_{mn} 
\nonumber\\
 &- (\hat{A} \leftrightarrow A)\,.
\end{eqnarray}
We will use a concise way of writing and in the following write only the coefficients of the terms, from which the term in question can be easily identified. The cancellation of the term (\ref{eq:term1}) gives the condition 
\begin{equation} \label{eq:cond1}
    1+\sum_{\alpha\beta=\{12,21\}}(-1)^\alpha C_{\alpha\beta112}=0\,.
\end{equation}
This is easily satisfied, however the terms in (\ref{eq:Term1}) will also give rise to other terms upon KK reduction. These terms must cancel out among themselves since there are no other terms of the same structure in (\ref{eq:tocancel}). In particular, taking the $n$-index to be internal gives the conditions
\begin{equation}\label{eq:cond2}
    \sum_{\alpha\beta=\{12,21\}}(-1)^\alpha C_{121\alpha\beta} = 0\quad \sum_{\alpha\beta=\{12,21\}}(-1)^\alpha C_{211\alpha\beta} = 0\,.
\end{equation}
This can again be satisfied by turning on more of the coefficients (those ending in $21$). But taking the $q$-index in these new terms to be internal gives the further condition
\begin{equation}
    \sum_{\alpha\beta=\{12,21\}}(-1)^\alpha C_{\alpha\beta121}=0\,,
\end{equation}
which, together with the previous two conditions, is incompatible with the first condition (\ref{eq:cond1}).

The conclusion is the same if we replace the metric by a vielbein, with KK reduction ansatz given in (\ref{eq:redf}). For the most general uplift term in (\ref{eq:Term1}) we now take
\begin{equation}
X^\alpha_{mn} = \{h_{mn}, B_{mn}, J_{mn}\}\quad\mbox{for}\quad\alpha = 1,2,3\,,
\end{equation}
where $h_{mn}=2f_{(mn)}$ and $J_{mn}=2f_{[mn]}$. 

Considering our previous term (\ref{eq:Term2}) in this new case, where now (by the $B$-parity) $C_{22112} = C_{11112} = C_{13112} = C_{31112} = C_{33112} = 0$ (i.e. odd in $2$'s), we get in place of condition (\ref{eq:cond1})
\begin{equation} \label{eq:cond1f}
    1+\sum_{\alpha\beta=\{12,21,23,32\}}(-1)^\alpha C_{\alpha\beta112}=0\,.
\end{equation}
Similar to the case for the linearized metric above one can now look at the terms coming from the  $n$-index being internal in each of the terms appearing in the sum and again find an inconsistency.

The same approach can be used to show that the other five terms in (\ref{eq:tocancel}) also cannot be uplifted to ten dimensions. It is clear what the problem is: any term in ten dimensions whose reduction cancels one of the terms in (\ref{eq:tocancel}) will also produce other terms, when another index is taken to be internal. But there are no other terms in (\ref{eq:tocancel}) with the same index structure which could cancel these terms.

So far we considered only the conventional fields and symmetries. One can ask whether approaches with additional fields and gauge symmetries, e.g. generalized geometry, might overcome the problem. But given how the problem arises it is hard to see how this could happen. In fact, we will see below that introducing additional vielbeins does not help.

\section{Double vielbein story}
Noting that the metric and the $B$ field each reduce to combinations of the KK vectors $A$ and $\hat A$, it is clear that to have the best chance of canceling the terms we want to cancel, we need objects which reduce just to $A$ or $\hat A$. This can be achieved by introducing two vielbeins instead of one, $e^{(\pm)}=1+f^{(\pm)}$ with (linearized) KK reduction ansatz
\begin{equation}
\underline{f}^{(+)}_{\underline{mn}} = \begin{pmatrix}
    f^{(+)}_{mn} & -A_{n'm} \\
    \hat{A}_{m'n} & 0
\end{pmatrix}\,, \quad \underline{f}^{(-)}_{\underline{mn}} = \begin{pmatrix}
    f^{(-)}_{mn} & \hat A_{n'm} \\
    - A_{m'n} & 0
\end{pmatrix}\,,
\end{equation}
hence
\begin{equation}
\underline{J}^{(\pm)}_{\underline{mn}} =\begin{pmatrix}
    J^{(\pm)}_{mn} &  \pm A^{(2)}_{n'm} \\
    \mp A^{(2)}_{m'n} & 0
\end{pmatrix}\,,
\end{equation}
where $h_{mn}=2f^{(\pm)}_{(mn)}$ and $J^{(\pm)}_{mn}=2f^{(\pm)}_{[mn]}$. Note that $f^{(+)}$ and $f^{(-)}$ are exchanged by $B$-parity.

We consider again the same term in (\ref{eq:term1}). For the general uplift term in (\ref{eq:Term1}) we now take
\begin{equation}
X^\alpha_{mn} = \{h_{mn}, B_{mn}, J^{(-)}_{mn},J^{(+)}_{mn}\}\quad\mbox{for}\quad\alpha = 1,2,3,4\,. 
\end{equation}
 In place of the condition (\ref{eq:cond1}) we now get
\begin{equation} \label{eq:cond1ff}
    1+\sum_{\alpha\beta=\{12,13,14,21,31,41\}}(-1)^{\alpha+\delta_{\beta3}}C_{\alpha\beta112}=0\,.
\end{equation}
Again each of the terms also give rise to other terms, by taking the $n$-index to be internal, which have to cancel separately. This leads to conditions similar to (\ref{eq:cond2}). Finally, by considering the $q$-index to be internal in the new terms one can derive the conditions
\begin{equation}
    \sum_{\gamma\delta=\{12,13,14,21,31,41\}}(-1)^{\gamma+\delta_{\delta3}}C_{\gamma\delta1\alpha\beta}=0
\end{equation}
and taking $\alpha\beta=12$ one finds again a contradiction with (\ref{eq:cond1ff}).

We conclude that introducing two independent vielbeins (and thereby doubling the Lorentz group) does not help in uplifting the terms in (\ref{eq:tocancel}) to ten dimensions. It seems very unlikely that introducing even more fields would change this conclusion. The reason is that any new field we introduce will still reduce to a combination of $A$ and $\hat A$ and their derivatives, but we can already produce any such combination using just $e^{(\pm)}$. This shows that approaches like generalized geometry or DFT/ExFT will not be able to account for the $\alpha'^3$ correction, as the $O(d,d)$ symmetry can only appear after KK reduction \emph{and} field redefinitions in the reduced theory.
\section{Conclusion and outlook}
It is known that at order $\alpha'$ and $\alpha'^2$ there is no obstruction to uplifting the field redefinitions needed for $O(d,d)$ invariance to $D=10$ (or $D=26$), which allows extended formulations with $O(d,d)$ symmetry before dimensional reduction \cite{Hohm:2014xsa,Marques:2015vua,Baron:2018lve,Baron:2020xel,Hronek:2021nqk, Hronek:2022dyr,Lunin:2024vsx}. Here we have argued that at order $\alpha'^3$ this is no longer possible~\footnote{Note that we are talking only about local field redefinitions. We have further assumed they do not explicitly break $10d$ Lorentz invariance by splitting the $10d$ index into subsets.}. This is consistent with the obstruction in DFT found in \cite{Hronek:2020xxi}. This presents a problem for formalisms with $O(d,d)$ built in, e.g. generalized geometry, when these are used in the uncompactified setting. This problem is present already for circle compactification ($d=1$), corresponding to $O(1,1)$ (it is however not visible for $d=9$, since the terms we considered vanish in that case). We have argued that it is unlikely to be overcome by introducing additional fields and gauge symmetries.

So far we did not discuss the possibility of modifying the $O(d,d)$ transformations. The idea would be to modify the $O(d,d)$ transformations at order $\alpha'^3$ so as to cancel the lowest order $O(d,d)$ \emph{transformation} of (\ref{eq:tocancel}), rather than the expression itself. For example, the first term implies a correction to the transformation of $B$ in the reduced theory of the form
\begin{equation}
\delta'B_{mn}\sim\alpha'^3\tilde\lambda^{a'b'}
(\hat F_{a'}F_{b'}\mathcal R\mathcal R)_{mn}\,,
\end{equation}
with $\tilde\lambda=\lambda-\hat\lambda$, the difference of the two $O(d)$ parameters inside $O(d,d)$. Naively, one can uplift this corrected transformation to a $D=10$ transformation of the form
\begin{equation}
\delta'B_{mn}\sim\alpha'^3\tilde\lambda^{ab}
(\omega_a H_b\mathcal R\mathcal R)_{mn}\,.
\end{equation}
Such a correction to the transformation does not seem to make much sense however, for the following reason. The combination of fields in the parenthesis cannot transform covariantly under (lowest order) $O(d,d)$. This means that the commutator of two $O(d,d)$ transformations will have terms of the form
\begin{equation}
[\delta_1,\delta_2]B_{mn}\sim\alpha'^3
(\omega\tilde\lambda_1H\tilde\lambda_2\mathcal R\mathcal R)_{mn}+\ldots-(1\leftrightarrow2)\,,
\end{equation}
i.e. with the two transformation parameters contracted with different fields, but this is not compatible with closure of the algebra. {This argument is only heuristic at this point, since it does not take into account all different ways one could implement the modification of the transformations, but it shows that there are serious problems that must be overcome to make sense of the required modifications to the transformations in $D=10$.}
For example, we expect this to be a problem for the approach of \cite{Baron:2022but}, which attempts to work in $D=10$ with the dimensional reduction treated implicitly (by dropping any terms with the $O(d,d)$ parameter contracted with a derivative).

On the positive side, we have seen that \emph{almost} all the required field redefinitions can be lifted to ten dimensions. The resulting Lagrangian (\ref{eq:L3}) is particularly natural to work with from a T-duality perspective. It would be important to extend it to all orders in the NSNS fields, and further to include also the RR fields of the type II string. This is particularly interesting as this correction plays an important role in areas such as black hole physics, e.g. \cite{Chen:2021qrz} 
, moduli stabilization, e.g. \cite{Burgess:2020qsc}, and inflation, e.g. \cite{Cicoli:2023njy, Cicoli:2024bwq}.

\section*{Acknowledgments}
We would like to thank Stanislav Hronek, Salomon Zacarias and Aviral Aggarwal for stimulating discussions and Arkady Tseytlin for comments on the draft. ARK thanks Michele Cicoli, Fernando Quevedo, Ashoke Sen and Ratul Mahanta for helpful discussions. This work is supported by the Czech Science Foundation (GAČR) grant ``Dualities and higher derivatives" (GA23-06498S).
\bibliographystyle{apsrev4-2}
\bibliography{biblio}{}
\appendix
\section{Kaluza-Klein reduction ansatz}
We use the standard KK reduction ansatz for the vielbein
\begin{equation}
\underline{e}_{\underline{m}}{}^{\underline{a}}=\left(\begin{array}{cc}
e_m{ }^a & A_m^{(1) n^{\prime}} e_{n^{\prime}}{ }^{a^{\prime}} \\
0 & e_{m^{\prime}}{ }^{a^{\prime}}
\end{array}\right)
\end{equation}
and $B$-field
\begin{equation}
\underline{B}_{\underline{m n}}=\left(\begin{array}{cc}
\tilde{B}_{m n} & A_{m n^{\prime}}^{(2)}+A_m^{(1) k^{\prime}} B_{k^{\prime} n^{\prime}} \\
-A_{n m^{\prime}}^{(2)}-A_n^{(1) k^{\prime}} B_{k^{\prime} m^{\prime}} & B_{m^{\prime} n^{\prime}}
\end{array}\right)\,,
\end{equation}
where $\tilde{B}_{m n}=B_{m n}-A_{[m}^{(1) m^{\prime}} A_{n] m^{\prime}}^{(2)}+A_m^{(1) k^{\prime}} A_n^{(1) l^{\prime}} B_{k^{\prime} l^{\prime}}$. The KK vectors can be combined into an $O(d,d)$ vector
\begin{equation}
\mathcal A_M=
\left(
\begin{array}{c}
	A^{(1)m'}\\
	A^{(2)}_{m'}
\end{array}
\right)
\end{equation}
and the scalars into a symmetric $O(d,d)$ matrix, sometimes called the generalized metric,
\begin{equation}
\mathcal H_{MN}=
\left(
\begin{array}{cc}
	g^{m'n'} & -g^{m'k'}B_{k'n'}\\
	B_{m'k'}g^{k'n'} & g_{m'n'}-B_{m'k'}g^{k'l'}B_{l'n'}
\end{array}
\right)\,.
\end{equation}
We also have the $O(d,d)$ invariant metric
\begin{equation}
\eta^{MN}=
\left(
\begin{array}{cc}
0 & \delta_{m'}{}^{n'}\\
\delta^{m'}{}_{n'} & 0	
\end{array}
\right)\,.
\end{equation}
Using these we can form precisely two $O(d,d)$ scalars from two KK vectors, namely
\begin{equation}
\mathcal A_M\eta^{MN}\mathcal A_N\qquad\mbox{and}\qquad\mathcal A_M\mathcal H^{MN}\mathcal A_N\,,
\end{equation}
where $\mathcal H^{MN}=\eta^{MK}\mathcal H_{KL}\eta^{LN}$. Since we will not need any terms involving the KK scalars we drop these in the following. In terms of $A$ and $\hat A$ defined in (\ref{eq:AAhat}) the two $O(d,d)$ invariants become
\begin{equation}
A_{m'm}A^{m'}_n\equiv A_mA_n\quad\mbox{and}\quad\hat A_{m'm}\hat A^{m'}_n\equiv \hat A_m\hat A_n\,,
\end{equation}
while
\begin{equation}
\hat A_{m'm}A^{m'}_n\equiv \hat A_mA_n
\end{equation}
violates $O(d,d)$ invariance.

The ``$B$-parity'' symmetry of the action under $\underline B\rightarrow-\underline B$ becomes $B\to -B$ and $A^{(2)} \to - A^{(2)}$ in the reduced theory, i.e.
\begin{equation}
    \hat{A} \to - A, \qquad A \to -\hat{A}
\end{equation}
which gives $\hat{A}_m A_n \to A_m \hat{A}_n$.

With the KK scalars set to zero the nontrivial contributions in the reduction of $H=dB$ become
\begin{equation}
\underline{H}_{a b c} = \tilde{H}_{a b c}\,, \quad \underline{H}_{a b c^{\prime}} =-F_{c^{\prime} a b}-\hat{F}_{c^{\prime} a b}\,,
\end{equation}
where $\tilde{H}_{a b c}=H_{a b c}+3\left(\hat{A}_{[a}\hat{F}_{b c]}-A_{[a}F_{b c]}\right)$ and $F$ and $\hat F$ are the field strengths of the KK vectors $A$ and $\hat A$ defined in (\ref{eq:AAhat}). The nontrivial contributions in the reduction of the torsionful Riemann tensor (\ref{eq:Rcal}) become
\begin{equation}
\begin{aligned}
\underline{\mathcal{R}}_{a b c^{\prime} d^{\prime}} &= -2 \hat{F}_{c^{\prime}[a}^e \hat{F}_{d^{\prime} b] e}\,, & 
\underline{\mathcal{R}}_{a b c^{\prime} d} &= \nabla_d^{(+)} \hat{F}_{c^{\prime} a b}\,, \\
\underline{\mathcal{R}}_{a^{\prime} b^{\prime} c d} &= -2 F_{a^{\prime}[c}^e F_{b^{\prime} d] e}\,, & 
\underline{\mathcal{R}}_{a^{\prime} b c d} &= -\nabla_b^{(-)} F_{a^{\prime} c d}\,, \\
\underline{\mathcal{R}}_{a^{\prime} b c^{\prime} d} &= \hat{F}_{c^{\prime} b}^e F_{a^{\prime} e d}\,, & 
\underline{\mathcal{R}}_{a b c d} &= \tilde{\mathcal{R}}_{a b c d} + \hat{F}_{a b}F_{c d}\,,
\end{aligned}
\end{equation}
where $\tilde{\mathcal{R}}_{a b c d} = \mathcal{R}_{a b c d} + 2 \hat{F}_{a[c}\hat{F}_{d] b} - F_{a b}F_{c d}$. Note that the last term in the reduction of $\mathcal R_{abcd}$ violates $O(d,d)$. 

The simplest example is the order $\alpha'^0$ action. Looking at the $O(d,d)$ violating terms one finds
\begin{equation}
\mathcal R\rightarrow-\hat F_{ab} F^{ab}\,,\qquad H^2\rightarrow 6\hat F_{ab}F^{ab}\,,
\end{equation}
fixing the correct relative coefficient in the Lagrangian, $L_{\alpha'^0}=\mathcal R+4(\partial\Phi)^2+\frac16H^2$. We refer to \cite{Wulff:2024ips} for further details. 
\section{Action and field redefinitions at order $\alpha'^3$}
We start from the action and $O(d,d)$ violating e.o.m. terms derived in \cite{Wulff:2024mgu}, eqs. (1.6) and (3.42). The only e.o.m. terms quartic in fields are contained in $k_5^{(\mathrm{e.o.m.})}$ in (B.43). Integrating these by parts one finds that they can be removed by adding to the ten-dimensional Lagrangian the following terms
\begin{equation}
\begin{split}
&\tfrac16(\varepsilon_7\varepsilon_7)'\mathcal R^3\hat{\mathcal R}
+\tfrac{1}{36}(\varepsilon_8\varepsilon_8)'H^2\mathcal R^2\hat{\mathcal R}
\\
&
-\tfrac16H^2(\varepsilon_5\varepsilon_5)'\mathcal R^2\hat{\mathcal R}
+\tfrac{1}{12}(\varepsilon_6\varepsilon_6)'\mathcal R^3\hat{\mathcal R}
\\
&
+\tfrac{1}{72}(\varepsilon_7\varepsilon_7)'H^2\mathcal R^2\hat{\mathcal R}
+\tfrac13H^2\mathcal R^{ab}{}_{cd}\mathcal R^{cd}{}_{ab}\hat{\mathcal R}\,,
\end{split}
\end{equation}
where the prime denotes removal of self-contractions (see \cite{Wulff:2024mgu}) and a hat denotes the completion of the torsionful Ricci tensor and scalar to the equations of motion
\begin{equation}
\begin{split}
\hat{\mathcal R}_{ab}&\equiv \mathcal R_{ab}+2\nabla^{(-)}_a\partial_b\Phi\,,\\
\hat{\mathcal R}&\equiv\mathcal R+4\nabla^d\partial_d\Phi-4\partial^d\Phi\partial_d\Phi+\tfrac16H^2\,,
\end{split}
\end{equation}
while $\hat{\mathcal R}^{ab}{}_{cd}\equiv\mathcal R^{ab}{}_{cd}$. This leaves the following $O(d,d)$ violating e.o.m. terms after KK reduction
\begin{equation}
\begin{split}
&
-\tfrac18(\varepsilon_7\varepsilon_7)'\hat FF\mathcal R^2\mathbb B
+4\hat F^{ab}F_{ab}\mathcal R^{cd}{}_{eg}\mathcal R^{ef}{}_{cd}\mathbb B^g{}_f
\\
&
+\tfrac12(\varepsilon_7\varepsilon_7)'\hat FF\nabla H\mathcal R\mathbb G
+\tfrac14(\varepsilon_6\varepsilon_6)'\hat FF\nabla H\mathcal R\mathbb D
\\
&
+\mathcal O(H^2)
+\mbox{(e.o.m.)$^2$-terms}
+(H\rightarrow-H, \hat F\leftrightarrow F)
\,.
\end{split}
\end{equation}
To this we have to add the remaining terms from (3.42) in \cite{Wulff:2024mgu}, which are all of fifth order in fields. Noting first that the contributions called $k_6^{(\mathrm{e.o.m.})}$ and $k_7^{(\mathrm{e.o.m.})}$ cancel out (to first order in $H$) one finds, after some algebra and integration by parts, that adding to the ten-dimensional Lagrangian the terms
\begin{equation}
\begin{split}
&\tfrac{5!^2}{2(3!)^2}H^{abc}H_{def}(\mathcal R^{[de}{}_{[ab}\mathcal R^{fg}{}_{cg}\hat{\mathcal R}^{h]}{}_{h]})'
\\
&+\tfrac{4!^2}{2(3!)^2}H^{abc}H_{def}(\mathcal R^{[de}{}_{[ab}\mathcal R^{fg]}{}_{cg]})'\hat{\mathcal R}
\\
&
+32H_a{}^{bc}H^{def}\mathcal R_{gh[de}\mathcal R^{ag}{}_{bf]}\hat{\mathcal R}^h{}_c
\\
&
+32H^a{}_{bc}H_{def}\mathcal R^{[de|gh|}\mathcal R^{bf]}{}_{ag}\hat{\mathcal R}^c{}_h
\\
&
-\tfrac{4!}{2}H^{abc}H^{def}\mathcal R^{gh}{}_{[ab}\mathcal R_{de]gh}\hat{\mathcal R}_{cf}\,,
\end{split}
\end{equation}
leaves only the $O(d,d)$ violating e.o.m. terms in (\ref{eq:tocancel}). (It is interesting to note that the second term in (\ref{eq:tocancel}) is special in that it arises only from the second term in the Lagrangian (\ref{eq:L3}), specifically from integrating the 3rd term in (3.7) in \cite{Wulff:2024mgu} by parts. This makes this term particularly easy to track.)

At the same time the ten-dimensional Lagrangian becomes as in (\ref{eq:L3}), where the various terms are defined as
\begin{equation}
\begin{split}
t_8t_8\mathcal R^4&=t_{a_1\cdots a_8}t^{b_1\cdots b_8}\mathcal R^{a_1a_2}{}_{b_1b_2}
\cdots\mathcal R^{a_7a_8}{}_{b_7b_8}
\\
t_8t_8H^2\mathcal R^3&=t_{a_1\cdots a_8}t^{b_1\cdots b_8}(H^2)^{a_1a_2}{}_{b_1b_2}
(\mathcal R^3)^{a_3\cdots a_8}{}_{b_3\cdots b_8}
\end{split}
\end{equation}
where $t_8$ is defined in the standard way
\begin{equation}
\begin{split}
t_8M_1M_2M_3&M_4
=
8\mathrm{tr}(M_1M_2M_3M_4)
\\
&
-2\mathrm{tr}(M_1M_2)\mathrm{tr}(M_3M_4)
+\mbox{cycl}(2,3,4)
\end{split}
\end{equation}
while
\begin{equation}
\begin{split}
\varepsilon_8\varepsilon_8\hat{\mathcal R}^4&=-8!\hat{\mathcal R}^{a_1a_2}{}_{[a_1a_2}
\cdots\hat{\mathcal R}^{a_7a_8}{}_{a_7a_8]}
\\
\varepsilon_9\varepsilon_9H^2\hat{\mathcal R}^3&=-9!H^{a_1a_2a_3}H_{[a_1a_2a_3}
(\hat{\mathcal R}^3)^{a_4\cdots a_9}{}_{a_4\cdots a_9]}
\end{split}
\end{equation}
and
\begin{equation}
\varepsilon_9\varepsilon_9[H^2]\hat{\mathcal R}^3=\tfrac{6!^2}{6}H^{a_1a_2a_3}H_{a_4a_5a_6}(\hat{\mathcal R}^3)^{[a_4\cdots a_9]}{}_{[a_1a_2a_3a_7a_8a_9]}
\end{equation}
which is simply the terms from the previous expression which have no contractions between the $H$'s. Finally we have defined the unconventional structure
\begin{equation}
\varepsilon_4H^2\varepsilon_4\mathcal R^2\mathcal R
=
-4!H_a{}^{a_1a_2}H^{ba_3a_4}\mathcal R_{ce[a_1a_2}\mathcal R_{a_3a_4]}{}^{ed}\mathcal R_{bd}{}^{ac}\,.
\end{equation}

\end{document}